# The use of Braid operators for implementing entangled large n-QUBITS Bell states (n>2)

Yacob Ben-Aryeh

Physics Department Technion-Israel Institute of Technology, Haifa, 32000, Israel

E-mail: phr65yb@physics.technion.ac.il

ABSTRACT

Braid theories are applied to quantum computation processes, where to each crossing in the Braid diagram a unitary $Yang-Baxter$ operator $R$ is associated, representing either a Braiding matrix or a universal quantum gate. By operating with Braid operators on the $computational$ basis of $n-qubits$ states, orthonormal entangled states are obtained, referred here as general Bell states. The $3-qubits$ Bell states are explicitly developed and the present methods are generalized to any $n-qubits$ system. The quantum properties of the general Bell states are analyzed and these properties are related to concurrence.

## 1. Introduction

In the present work Braids theories [1-2] are applied showing how to implement Bell entanglement in large $n-qubits$ systems (n>2) [3-4]. Braids form a group under concatenation. In concatenation the bottom strands of the first braid are attached to the top strands of the second braid. A fundamental concept in the present use of the Braid representation is the association of Yang-Baxter operator $R$ [5-7] to each elementary crossing in the braid diagram. Such operator is not necessarily unitary in topological applications [1-2], but for the purpose of quantum computation (QC) we restrict the operator $R$ to be unitary. Then, the $R$ matrix represents either a braiding matrix or a quantum gate in QC.

Any quantum gate can be given by combinations of a universal gate representing entangling processes and local single $qubit$ transformations [8]. Such universal gate is well known as the CNOT gate [3-4] but there are other universal gates which can be related to the CNOT gate by local single *qubit* transformations. In the present analysis we will use the universal $R$ gate [5-6], representing also a special unitary solution to the Yang-Baxter equation [5-7]. We take into account that CNOT gate does not satisfy the Yang-Baxter equation

We consider the projection of Braids in the plane as a set of n strings all of which are attached to an upper horizontal line leading downwards to a bottom horizontal line. As we are interested in the implementation of Braids theories to QC, equally spaced points ("bullets") are described in the present schemes on the upper line denoted as $1, 2, \cdots, i, i+1, i+2, \cdots, n$ where each point represents a *qubit*. The strands starting at the points in the upper line represent the initial states of *qubits* and they are going downwards representing their time development. The number n represents the number of *qubits* included in our system. Each strand can pass a neighboring strand at a certain point where in our scheme such "crossing" represents unitary interaction between consequent *qubits*. In each such crossing the solid line in the "crossing" is assumed to move somewhat above the plane while the line which is hollow (has a break) near the crossing is assumed to move somewhat below the crossing. The crossings between *qubits i* and $i+1$ represent special unitary transformations operating on the two *qubits* given by the $R$ matrix (or $R^{-1}$). On each horizontal line between the upper and lower horizontal lines there is only one crossing. The strands are ending finally in our schemes at equally spaced points in the lower line representing the final states of the *qubits*.

In the present paper we use a special representation for the *Bn Artin* group [2]. We develop $n-qubits\ states$ in time by the Braid operators, operating on the pairs $(1,2), (2,3), (3,4)$ etc., consequently. Each pair interaction is given by the $R$ matrix satisfying also the Yang-Baxter equation. By performing such unitary interactions on a *computational* basis of $n-qubits$ states [3-4], we will get $n-qubits$ entangled orthonormal Bell basis of states. The properties of the Bell states which generalize the properties of the $2-qubits$ Bell states to $n-qubits$ states ($n$ >2), are analyzed in the present work. The $3-qubits$ Bell entangled states are developed explicitly and we show how to extend the calculations for any $n-qubits$ Bell entangled states. The entanglement properties are related to concurrence [9-12].

The present paper is organized as follows:

In Section 2, I describe the special representation for the Braid operators satisfying $B_n$ Artin group [2] relations, and correspondingly describe the universal $R$ gate, satisfying the Yang-Baxter equation [5-6]. In Section 3, certain analogies between Braid diagrams, satisfying the Yang-Baxter equation, and QC, are described. In Section 4, I develop the use of Braid operators for implementing the entangled $n-qubits$ states, referred here as general Bell states. We analyze especially the $3-qubits$ Bell entangled states. In Section 5 the entanglement properties of the general Bell states will be related to concurrence and entanglement of information [9-12]. The properties of the $3-qubits$ Bell states are compared with those of the GHZ states [13-14]. In Section 6, we discuss the present results, and summarize our results and conclusions.

## 2. Special representation for the $B_n$ Braid group

The $R$ matrix is given in the present work by [6]:

$$R = \frac{1}{\sqrt{2}} \begin{pmatrix} 1 & 0 & 0 & 1 \\ 0 & 1 & -1 & 0 \\ 0 & 1 & 1 & 0 \\ -1 & 0 & 0 & 1 \end{pmatrix}. \qquad (1)$$

The $computational$ basis of $2-qubits$ states [3-4] is given by:

$$|C1\rangle = |0\rangle_A |0\rangle_B = \begin{pmatrix} 1 \\ 0 \end{pmatrix} \otimes \begin{pmatrix} 1 \\ 0 \end{pmatrix} = \begin{pmatrix} 1 \\ 0 \\ 0 \\ 0 \end{pmatrix} \;;\; |C2\rangle = |0\rangle_A |1\rangle_B = \begin{pmatrix} 1 \\ 0 \end{pmatrix} \otimes \begin{pmatrix} 0 \\ 1 \end{pmatrix} = \begin{pmatrix} 0 \\ 1 \\ 0 \\ 0 \end{pmatrix} \;;$$

$$|C3\rangle = |1\rangle_A |0\rangle_B = \begin{pmatrix} 0 \\ 1 \end{pmatrix} \otimes \begin{pmatrix} 1 \\ 0 \end{pmatrix} = \begin{pmatrix} 0 \\ 0 \\ 1 \\ 0 \end{pmatrix} \;;\; |C4\rangle = |1\rangle_A |1\rangle_B = \begin{pmatrix} 0 \\ 1 \end{pmatrix} \otimes \begin{pmatrix} 0 \\ 1 \end{pmatrix} = \begin{pmatrix} 0 \\ 0 \\ 0 \\ 1 \end{pmatrix} \;;$$

(2)

Here, $|0\rangle = \begin{pmatrix} 1 \\ 0 \end{pmatrix}$ and $|1\rangle = \begin{pmatrix} 0 \\ 1 \end{pmatrix}$ are the two states of each *qubit* and the subscripts $A, B$ refer to the first and second *qubit*, respectively. The notations $|C1\rangle, |C2\rangle |C3\rangle |C4\rangle$ refer to the *computational* basis of states where these states can be described by 4-dimensional vectors. It is quite straightforward to find that by operating with the $R$ matrix on the 4-dimensional vectors of (2) we get the $2-qubits$ Bell entangled states. We are interested, however, in implementation of Braid theory to entanglement in large $n-qubits$ system.

Equation (1) satisfy also a special Yang-Baxter equation [5-6] given by

$$(R \otimes I) \cdot (I \otimes R) \cdot (R \otimes I) = (I \otimes R) \cdot (R \otimes I) \cdot (I \otimes R) \quad . \tag{3}$$

Here the dot represents ordinary matrix multiplication, the symbol $\otimes$ represents outer product, $I$ is unit $2 \times 2$ matrix, and where $(R \otimes I)$ and $(I \otimes R)$ are matrices of $8 \times 8$ dimensions. The unitary $R$ matrix can be considered in QC as a universal gate [8], so that all entanglement processes in QC can be described by the $R$ matrix satisfying the Yang Baxter equation, plus the use of local single *qubit* transformations.

We use for our system the finite representation discovered by *Artin* [2], where the generators $\sigma_1, \sigma_2, \cdots, \sigma_{n-1}$ satisfy the *Bn* group relations:

$$B_n = \left\langle \begin{array}{l} \sigma_1, \sigma_2, \cdots, \sigma_{n-1} \mid \sigma_i \sigma_j = \sigma_j \sigma_i \quad |i-j| > 1 \; ; \\ \sigma_i \sigma_j \sigma_i = \sigma_j \sigma_i \sigma_j \quad |i-j| = 1 \; ; \quad \sigma_i \sigma_i^{-1} = I \end{array} \right\rangle \quad . \tag{4}$$

For the *Bn* operators $\sigma_1, \sigma_2, \cdots, \sigma_{n-1}$, operating in our system on the $n-qubit$ states, we use the representation:

$$\begin{aligned}
\sigma_1 &= R \otimes I \otimes I \cdots \otimes I \; ; \\
\sigma_2 &= I \otimes R \otimes I \cdots \otimes I \; ; \\
\sigma_2 &= I \otimes I \otimes R \cdots \otimes I \; ; \\
&\cdot \\
&\cdot \\
&\cdot \\
\sigma_{n-1} &= I \otimes I \otimes I \cdots \otimes I \otimes R
\end{aligned} \tag{5}$$

Here $R$ is the unitary matrix given by (1) and $I$ is the unit $2 \times 2$ matrix.

One should take into account that the operators $R$ and $I$ on the right side of (5) operate in the Braid diagram consequently on the *qubits* $1, 2, \cdots, n$, where $R$ entangles two consequent *qubits* and the $I$ matrices represent *qubits* which are non-interacting. One should notice also that for $\sigma_1, \sigma_2, \cdots, \sigma_{n-1}$ the entanglement is produced between *qubits* $(1,2), (2,3), (3,4) \cdots (n-1, n)$, correspondingly. It is quite easy to verify that the $Bn$ group relations (4) are satisfied, using (3) and (5).

We will develop in the next Section certain analogs between Braid diagrams and QC, related to the use of Yang-Baxter equation.

## 3. Quantum entanglement in a $3-qubits$ system, described by a Braid scheme satisfying Yang-Baxter Equation

By using (5) for a $3-qubits$ system we get the $Bn$ operators

$$\sigma_1 = R \otimes I \quad ; \quad \sigma_2 = I \otimes R \quad . \tag{6}$$

These operators satisfy the relations

$$\sigma_1 \sigma_2 \sigma_1 = \sigma_1 \sigma_2 \sigma_1 \quad . \tag{7}$$

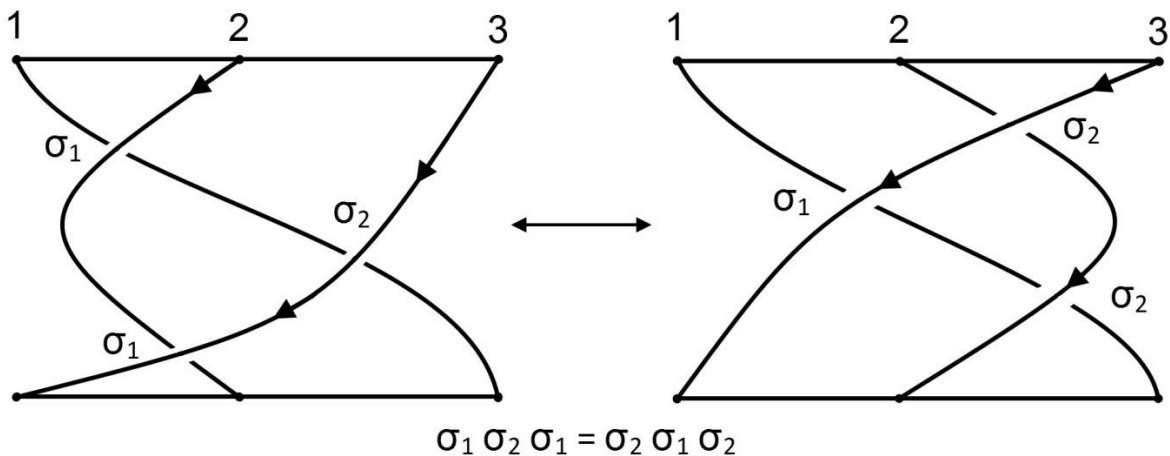

Figure1:

The interactions between the three *qubits* 1,2, and 3 is shown to satisfy the Yang-Baxter equation where the Braid operators $\sigma_1$ and $\sigma_2$ are represented by $R \otimes I$ and $I \otimes R$, respectively, $R$ is given by (1), and $I$ is the $2 \times 2$ unit matrix. The $8 \times 8$ matrix $\sigma_1$ ($\sigma_2$) entangles the *qubits* 1 and 2 (2 and 3).

This equation is equivalent to the Yang-Baxter equation (3). The quantum entanglement in a $3-qubits$ system satisfying Yang-Baxter equation is described by the Braid diagrams in *Figure* 1 Crossings in this *Figure* represent entangling interactions which are given by $\sigma_i$ $(i=1,2)$. A movement above the crossing, of the right hand from the solid line denoted with an arrow to the other hollow line at the crossing is in opposite clock direction. If such movement will be in the clock direction then the crossing will represent $\sigma_i^{-1}$ (See a demonstration of such crossing in Figure 2).

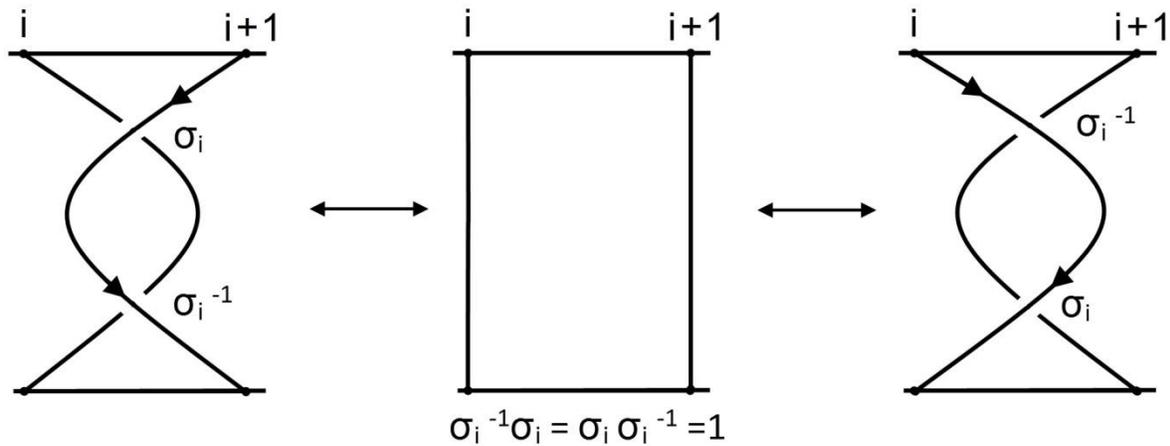

$$\sigma_i^{-1}\sigma_i = \sigma_i \sigma_i^{-1} = 1$$

Figure 2

A crossing represents the Braid operator $\sigma_i \left(\sigma_i^{-1}\right)$, where a movement, above the crossing, of the right hand from the solid line denoted with an arrow to the other hollow line of the crossing is in opposite clock direction (in the clock direction). $i$, $i+1$ represent two neighboring *qubits* in which the inverse interactions are cancelled, and which are equivalent to *qubits* which are not interacting, represented by the vertical lines.

Two Braids are equivalent if one can be transformed to the other by combinations of Re*idemeister* moves (see e.g. [15]). There is a certain analogy between the Re*idemeister* moves and the use of the Yang-Baxter equation. One can notice that the curves on the right side of *Figure* 1 corresponding to the transformation $\sigma_2\sigma_1\sigma_2$ and the curves on the left side corresponding to $\sigma_1\sigma_2\sigma_1$ are symmetric relative to the vertical line passing through the $i+1$ point. Such symmetry follows from the property of Yang-Baxter equation and is similar to the symmetry obtained in Braids theory by the third Re*idemeister* move [15]. The second

Re*idemeister* move allows us to remove two special crossing points. In QC there is an analogous trivial effect where entanglement produced by the matrix $\sigma_i$ can be removed by the inverse operation $\sigma^{-1}$. Such effect is described in *Figure* 2. The vertical lines in this *Figure* represent non-interacting single *qubits* where the entanglement between two neighboring *qubits* has been eliminated by the inverse unitary transformation. In Braid theory the first Re*idemeister* move allows us to put in, or take out, a twist in the Braid strands. In QC the analog to the first Re*idemeister* move, will be the local single *qubits* transformation. While in Braid theories the first Re*idemeister* move has a simple effect, in QC the local single *qubit* transformations can become quite complicated and only by involving such transformations with the entangling processes, we can analyze QC effects. On the other hand the $\sigma_i$ operators in topological applications are not restricted to be unitary.

**4. General Entangled Bell-states related to the $\sigma_i$ operators**

We develop first the general Bell entangled states for a $3-qubits$ system. Afterwards we will show how to generalize the present methods to any $n-qubit$ system.

We will denote again, $|0\rangle = \begin{pmatrix} 1 \\ 0 \end{pmatrix}$ and $|1\rangle = \begin{pmatrix} 0 \\ 1 \end{pmatrix}$ as the two states of each *qubit* and the subscripts $A, B$ and $C$ will refer to the first, second and third *qubit*, respectively. The notations $|C1\rangle, |C2\rangle \cdots |C8\rangle$ will refer to the *computational* basis of states where these states can be described by 8-dimensional vectors. In the $i'th$ entry of the vector $|Ci\rangle$ $(i = 1, 2, \cdots, 8)$ one gets $1$, and in all other entries one gets $0$.

The *computational* basis of $3-qubits$ states [3-4] is given by the following equivalent forms:

$$|C1\rangle = |0\rangle_A |0\rangle_B |0\rangle_C = \begin{pmatrix} 1 \\ 0 \end{pmatrix} \otimes \begin{pmatrix} 1 \\ 0 \end{pmatrix} \otimes \begin{pmatrix} 1 \\ 0 \end{pmatrix} = \begin{pmatrix} 1 \\ 0 \\ 0 \\ 0 \\ 0 \\ 0 \\ 0 \\ 0 \end{pmatrix} \quad ; \quad |C2\rangle = |0\rangle_A |0\rangle_B |1\rangle_C = \begin{pmatrix} 1 \\ 0 \end{pmatrix} \otimes \begin{pmatrix} 1 \\ 0 \end{pmatrix} \otimes \begin{pmatrix} 0 \\ 1 \end{pmatrix} = \begin{pmatrix} 0 \\ 1 \\ 0 \\ 0 \\ 0 \\ 0 \\ 0 \\ 0 \end{pmatrix} \quad ;$$

$$|C3\rangle = |0\rangle_A |1\rangle_B |0\rangle_C = \begin{pmatrix} 1 \\ 0 \end{pmatrix} \otimes \begin{pmatrix} 0 \\ 1 \end{pmatrix} \otimes \begin{pmatrix} 1 \\ 0 \end{pmatrix} = \begin{pmatrix} 0 \\ 0 \\ 1 \\ 0 \\ 0 \\ 0 \\ 0 \\ 0 \end{pmatrix} \quad ; \quad |C4\rangle = |0\rangle_A |1\rangle_B |1\rangle_C = \begin{pmatrix} 1 \\ 0 \end{pmatrix} \otimes \begin{pmatrix} 0 \\ 1 \end{pmatrix} \otimes \begin{pmatrix} 0 \\ 1 \end{pmatrix} = \begin{pmatrix} 0 \\ 0 \\ 0 \\ 1 \\ 0 \\ 0 \\ 0 \\ 0 \end{pmatrix} \quad ;$$

$$|C5\rangle = |1\rangle_A |0\rangle_B |0\rangle_C = \begin{pmatrix} 0 \\ 1 \end{pmatrix} \otimes \begin{pmatrix} 1 \\ 0 \end{pmatrix} \otimes \begin{pmatrix} 1 \\ 0 \end{pmatrix} = \begin{pmatrix} 0 \\ 0 \\ 0 \\ 0 \\ 1 \\ 0 \\ 0 \\ 0 \end{pmatrix} \quad ; \quad |C6\rangle = |1\rangle_A |0\rangle_B |1\rangle_C = \begin{pmatrix} 0 \\ 1 \end{pmatrix} \otimes \begin{pmatrix} 1 \\ 0 \end{pmatrix} \otimes \begin{pmatrix} 0 \\ 1 \end{pmatrix} = \begin{pmatrix} 0 \\ 0 \\ 0 \\ 0 \\ 0 \\ 1 \\ 0 \\ 0 \end{pmatrix} \quad ;$$

$$|C7\rangle = |1\rangle_A |1\rangle_B |0\rangle_C = \begin{pmatrix} 0 \\ 1 \end{pmatrix} \otimes \begin{pmatrix} 0 \\ 1 \end{pmatrix} \otimes \begin{pmatrix} 1 \\ 0 \end{pmatrix} = \begin{pmatrix} 0 \\ 0 \\ 0 \\ 0 \\ 0 \\ 0 \\ 1 \\ 0 \end{pmatrix} ; \quad |C8\rangle = |1\rangle_A |1\rangle_B |1\rangle_C = \begin{pmatrix} 0 \\ 1 \end{pmatrix} \otimes \begin{pmatrix} 0 \\ 1 \end{pmatrix} \otimes \begin{pmatrix} 0 \\ 1 \end{pmatrix} = \begin{pmatrix} 0 \\ 0 \\ 0 \\ 0 \\ 0 \\ 0 \\ 0 \\ 1 \end{pmatrix} .$$

(8) .

The present entangled states for a $3-qubits$ system are obtained by operating on the states vectors of (8) with $\sigma_1 \cdot \sigma_2 = (R \otimes I) \cdot (I \otimes R)$. A straight forward calculation for the multiplication $\sigma_1 \cdot \sigma_2 = (R \otimes I) \cdot (I \otimes R)$ gives the result:

$$\sigma_1 \cdot \sigma_2 = \begin{pmatrix} 1 & 0 & 0 & 1 & 0 & 1 & 1 & 0 \\ 0 & 1 & -1 & 0 & -1 & 0 & 0 & 1 \\ 0 & 1 & 1 & 0 & -1 & 0 & 0 & -1 \\ -1 & 0 & 0 & 1 & 0 & -1 & 1 & 0 \\ 0 & 1 & 1 & 0 & 1 & 0 & 0 & 1 \\ -1 & 0 & 0 & 1 & 0 & 1 & -1 & 0 \\ -1 & 0 & 0 & -1 & 0 & 1 & 1 & 0 \\ 0 & -1 & 1 & 0 & -1 & 0 & 0 & 1 \end{pmatrix}. \quad (9)$$

By operating with the matrix $\sigma_1 \cdot \sigma_2$ on the computational basis of states vectors (8) we get the corresponding Bell entangled basis of states for $3-qubits$ system. The Bell entangled basis of states are then given by superposition of computational states and it is convenient to write them in the following form:

$$\begin{aligned}
2|B1\rangle &\equiv |0\rangle_A |0\rangle_B |0\rangle_C - |0\rangle_A |1\rangle_B |1\rangle_C - |1\rangle_A |0\rangle_B |1\rangle_C - |1\rangle_A |1\rangle_B |0\rangle_C \;; \\
2|B2\rangle &\equiv |0\rangle_A |0\rangle_B |1\rangle_C + |0\rangle_A |1\rangle_B |0\rangle_C + |1\rangle_A |0\rangle_B |0\rangle_C - |1\rangle_A |1\rangle_B |1\rangle_C \;; \\
2|B3\rangle &\equiv -|0\rangle_A |0\rangle_B |1\rangle_C + |0\rangle_A |1\rangle_B |0\rangle_C + |1\rangle_A |0\rangle_B |0\rangle_C + |1\rangle_A |1\rangle_B |1\rangle_C \;; \\
2|B4\rangle &\equiv |0\rangle_A |0\rangle_B |0\rangle_C + |0\rangle_A |1\rangle_B |1\rangle_C + |1\rangle_A |0\rangle_B |1\rangle_C - |1\rangle_A |1\rangle_B |0\rangle_C \;; \\
2|B5\rangle &\equiv -|0\rangle_A |0\rangle_B |1\rangle_C - |0\rangle_A |1\rangle_B |0\rangle_C + |1\rangle_A |0\rangle_B |0\rangle_C - |1\rangle_A |1\rangle_B |1\rangle_C \;; \\
2|B6\rangle &\equiv |0\rangle_A |0\rangle_B |0\rangle_C - |0\rangle_A |1\rangle_B |1\rangle_C + |1\rangle_A |0\rangle_B |1\rangle_C + |1\rangle_A |1\rangle_B |0\rangle_C \;; \\
2|B7\rangle &\equiv |0\rangle_A |0\rangle_B |0\rangle_C + |0\rangle_A |1\rangle_B |1\rangle_C - |1\rangle_A |0\rangle_B |1\rangle_C + |1\rangle_A |1\rangle_B |0\rangle_C \;; \\
2|B8\rangle &\equiv |0\rangle_A |0\rangle_B |1\rangle_C - |0\rangle_A |1\rangle_B |0\rangle_C + |1\rangle_A |0\rangle_B |0\rangle_C + |1\rangle_A |1\rangle_B |1\rangle_C \;.
\end{aligned} \quad (10)$$

One can easily check that the quantum states $|B1\rangle, |B2\rangle, \cdots, |B8\rangle$ form another orthonormal basis of states for the $3-qubits$ system which has special entanglement properties. Any one of the $3-qubits$ Bell states includes superposition of 4 multiplications with equal probability (i.e., real amplitudes are either 1 or -1), and each state in the same multiplication belongs to a different *qubit*. This property is analogous to the $2-qubits$ Bell states property, where we have two multiplications with equal probability and the first and second state in each Bell state belong to the first and second *qubit*, respectively.

The method analyzed above can be generalized for implementing any $n-qubits$ Bell states by operating with the matrix $\sigma_1 \cdot \sigma_2 \cdots \sigma_{n-1}$ on the *computational* states $|C1\rangle, |C2\rangle, \cdots, |C2^n\rangle$. Any Bell state obtained by this method will include $2^{n-1}$ multiplications, with equal probability, where each state in the same multiplication belongs to another *qubit*, from the different $n$ *qubits*. In our previous examples we found for the $2-qubits$ Bell states $2^{n-1} = 2$ multiplications, and for the $3-qubits$ $2^{n-1} = 4$ multiplications. The application of the method for any $n-qubits$ system is straight forward but the computation becomes quite tedious for a system with a large number of *qubits* representing the entanglement complexity.

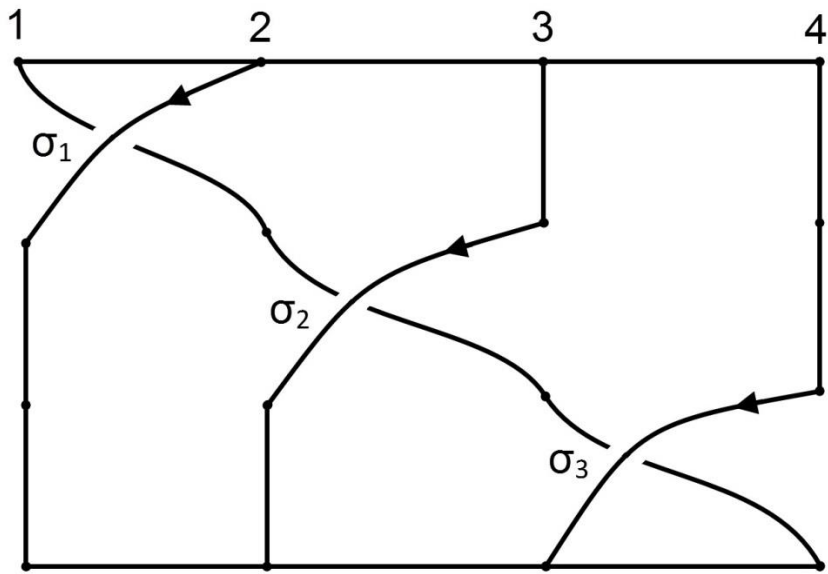

Figure 3

The numbers, on the upper horizontal line, denote *qubits* $1, 2, 3, 4$, respectively. The crossings represent the Braid operators $\sigma_1, \sigma_2$ and $\sigma_3$, given by (5), entangling the *qubits* pairs $(1,2), (2,3), (3,4)$, consequently. Solid lines connecting neighboring points on vertical lines represent single *qubits* which are not interacting.

The Braid diagram for the braid operator $\sigma_1 \cdot \sigma_2 \cdot \sigma_3$ which can lead to Bell entangled $4-qubits$ is described schematically in $Figure\ 3$. Any Bell $4-qubits$ state obtained by operating with $\sigma_1 \cdot \sigma_2 \cdot \sigma_3$ on any state from the 16 $computational$ states, will include superposition of $2^{n-1}=8$ multiplications with equal probabilities, in which each state in the same multiplication belongs to a different $qubit$ from the 4 different $qubits$. It is straightforward to find the $16\times 16$ matrix $\sigma_1 \cdot \sigma_2 \cdot \sigma_3$ and the complete orthonormal basis of $4-qubits$ Bell states following the above method. For the simplicity of presentation I present here only the $4-qubits$ state $|B'1\rangle$ which is obtained by operating with $\sigma_1 \cdot \sigma_2 \cdot \sigma_3$ on the computational 16 dimensional $|C'1\rangle$ vector which has 1 in the first entry and zero in all other entries. I get:

$$|B'1\rangle = |0\rangle_A |0\rangle_B |0\rangle_C |0\rangle_D - |0\rangle_A |0\rangle_B |1\rangle_C |1\rangle_D - |0\rangle_A |1\rangle_B |0\rangle_C |1\rangle_D - |0\rangle_A |1\rangle_B |1\rangle_C |0\rangle_D$$
$$- |1\rangle_A |0\rangle_B |0\rangle_C |1\rangle_D - |1\rangle_A |0\rangle_B |1\rangle_C |0\rangle_D - |1\rangle_A |1\rangle_B |0\rangle_C |0\rangle_D + |1\rangle_A |1\rangle_B |1\rangle_C |1\rangle_D \quad (11)$$

The entanglement obtained in the general Bell states will be related in the next Section to concurrence and measurement properties.

## 5. Entanglement properties of the general Bell states related to concurrence and to separability

Let us consider first a pure state $|\Phi\rangle$ of $qubits$ pair. The concurrence $C(|\Phi\rangle)$ of this state [9-12] is defined to be

$$C(|\Phi\rangle) = |\langle \Phi | \tilde{\Phi} \rangle| \quad . \quad (12)$$

Here the tilde denotes spin flip operation [9]:

$$|\tilde{\Phi}\rangle = \sigma_y \otimes \sigma_y |\Phi^*\rangle \quad , \quad (13)$$

$|\Phi^*\rangle$ is the complex conjugate of $|\Phi\rangle$, in the standard basis $\{|00\rangle, |01\rangle, |10\rangle, |11\rangle\}$, and $\sigma_y$ is the Pauli spin operator $\begin{pmatrix} 0 & -i \\ i & 0 \end{pmatrix}$. The spin flip operation takes the state of each $qubit$ in a pure

product state to the orthogonal state so that the concurrence of a pure product state is zero. On the other hand a completely entangled state such as $2-qubits$ Bell state is invariant under spin flip so that for such state the concurrence $C(|\Phi\rangle)$ takes the value $1$ which is the maximal possible value of $C$.

For the more general case in which the density matrix can be mixed, the "spin flip" density matrix is defined to be [9-10]

$$\tilde{\rho}_{A,B} = (\sigma_y \otimes \sigma_y) \rho^*_{A,B} (\sigma_y \otimes \sigma_y) \qquad . \tag{14}$$

Here the asterisk denotes complex conjugation. The product $\rho_{A,B}\tilde{\rho}_{A,B}$ has only real and non-negative values and the square roots of these eigenvalues in decreasing order are $\lambda_1, \lambda_2, \lambda_3, \lambda_4$. Then, the concurrence of the density matrix $\rho_{A,B}$ is defined as

$$C_{AB} = \max\{\lambda_1 - \lambda_2 - \lambda_3 - \lambda_4, 0\} \tag{15}$$

We find the interesting point that by taking the density matrix of any $3-qubits$ Bell $|Bi\rangle\langle Bi|$ $(i=1,2,\cdots,8)$ entangled state and by tracing over any single $qubit$ ($A$, or $B$, or $C$) the mixed $2-qubits$ states is obtained with concurrence equal to zero. It quite easy to verify this result by making the calculation for a typical example given as

$$4\rho_{A,B} = 4Tr_C|B1\rangle\langle B1| = (|0\rangle_A|0\rangle_B - |1\rangle_A|1\rangle_B)(\langle 0|_A\langle 0|_B - \langle 1|_A\langle 1|_B) \\ + (|0\rangle_A|1\rangle_B + |1\rangle_A|0\rangle_B)(\langle 1|_A\langle 0|_B + \langle 0|_A\langle 1|_B) \qquad . \tag{16}$$

In the standard basis this density matrix can be written as

$$4\rho_{A,B} = \begin{pmatrix} 1 & 0 & 0 & -1 \\ 0 & 1 & 1 & 0 \\ 0 & 1 & 1 & 0 \\ -1 & 0 & 0 & 1 \end{pmatrix} = 4\rho^*_{A,B} \qquad . \tag{17}$$

The product $\rho_{A,B}\tilde{\rho}_{A,B}$ is calculated and given by

$$\rho_{A,B}\tilde{\rho}_{A,B} = \rho_{A,B}\left(\sigma_y \otimes \sigma_y\right)\rho^*_{A,B}\left(\sigma_y \otimes \sigma_y\right) =$$

$$\frac{1}{16}\begin{pmatrix} 1 & 0 & 0 & -1 \\ 0 & 1 & 1 & 0 \\ 0 & 1 & 1 & 0 \\ -1 & 0 & 0 & 1 \end{pmatrix}\begin{pmatrix} 0 & 0 & 0 & -1 \\ 0 & 0 & 1 & 0 \\ 0 & 1 & 0 & 0 \\ -1 & 0 & 0 & 0 \end{pmatrix}\begin{pmatrix} 1 & 0 & 0 & -1 \\ 0 & 1 & 1 & 0 \\ 0 & 1 & 1 & 0 \\ -1 & 0 & 0 & 1 \end{pmatrix}\begin{pmatrix} 0 & 0 & 0 & -1 \\ 0 & 0 & 1 & 0 \\ 0 & 1 & 0 & 0 \\ -1 & 0 & 0 & 0 \end{pmatrix} \quad (18)$$

$$= \frac{1}{8}\begin{pmatrix} 1 & 0 & 0 & -1 \\ 0 & 1 & 1 & 0 \\ 0 & 1 & 1 & 0 \\ -1 & 0 & 0 & 1 \end{pmatrix}$$

It is quite easy to find that the square roots of the eigenvalues of (18) are given by $\lambda_1 = \lambda_2 = \frac{1}{2}$ ; $\lambda_3 = \lambda_4 = 0$ so that according to (15) the concurrence is equal to $0$, which means that after tracing over the $qubit$ $C$ we remain with the $2-qubits$ mixed state which has only classical correlations. The same conclusion is obtained by using *Peres-Horodecki* criterion [16-17]. Following this criterion we partly transpose (PT) the density matrix $\rho_{A,B}$ for the $qubit$ B as $|k\rangle_B \langle l|_B \to |l\rangle_B \langle k|_B$ ; $l,k = 1,2$ and leave the $qubit$ A unchanged. Then the PT of $\rho_{A,B}$ is given by

$$4\rho_{A,B}(PT) = |0\rangle_A|0\rangle_B\langle 0|_A\langle 0|_B + |1\rangle_A|1\rangle_B\langle 1|_A\langle 1|_B - |1\rangle_A|0\rangle_B\langle 0|_A\langle 1|_B - |0\rangle_A|1\rangle_B\langle 1|_A\langle 0|_B$$
$$|1\rangle_A|0\rangle_B\langle 1|_A\langle 0|_B + |0\rangle_A\langle 1|_B\langle 0|_A\langle 1|_B + |1\rangle_A|1\rangle_B\langle 0|_A\langle 0|_B + |0\rangle_A|0\rangle_B\langle 1|_A\langle 1|_B \quad (19)$$

In the standard basis $4\rho_{A,B}(PT)$ can be written as

$$4\rho_{A,B}(PT) = \begin{pmatrix} 1 & 0 & 0 & 1 \\ 0 & 1 & -1 & 0 \\ 0 & -1 & 1 & 0 \\ 1 & 0 & 0 & 1 \end{pmatrix} \quad (20)$$

The eigenvalues of $\rho_{A,B}(PT)$ are $1/2;1/2;0;0$. As they are non-negative $\rho_{A,B}$ is 'separable' so that it includes only classical correlations.

The result obtained here by which the density matrix (16) is separable with zero concurrence seems to be quite interesting as this density matrix is composed of a superposition of two $2-qubits$ Bell density matrices, where each of them has a maximal entanglement (i.e. concurrence equal 1). However, such entanglement is cancelled as in the calculation of the concurrence we obtained the square roots $\lambda_1 = \lambda_2 \, ; \lambda_3 = \lambda_4 = 0$ and concurrence vanishes according to (15). Also the density matrix becomes separable following Peres-Horodecki criterion. We find that the $3-qubits$ Bell state has a maximal entanglement property but in order to realize such entanglement we need $3-qubits$ measurements. If we trace over a $qubit$, i.e. ignore its measurement, the quantum entanglement is cancelled.

Although we have made explicit calculation for one example, due to symmetry properties of the entangled states of (10) we get the following general conclusion: By tracing *any qubit* from *any Bell entangled* $3-qubit$ state given by (10) we get a mixed state which has zero concurrence and separable density matrix, so that it includes only classical correlations. In a pictorial description: Assuming that we have *Bell entangled* $3-qubit$ state and we send the three *qubits* to Alice, Bob and Charles, respectively, which are far, each from the other. If we ignore any measurement made by one of them (e.g. by Charles) then the other two (e.g. Alice and Bob) can have only classical correlations between them.

The above properties of having $3-qubits$ entangled states, where by tracing over one *qubit* a mixed state is obtained which is separable with zero concurrence, are obtained also for the GHZ states [13-14]. There are, however, special properties for the present Bell entangled states: 1) The Bell entangled states form a complete set of orthonormal states for any $n-qubit$ system. 2) The implementation of these states is based on the operation of the Braid operator multiplication $\sigma_1 \sigma_2 \cdots \sigma_{n-1}$. Although there is an enormous amount of literature showing how to implement universal gates, including the CNOT gate (see e.g. [18]) the use of the Braid method should use a special technique. We should notice that the $n-qubit$ entangled Bell states can be produced by entangling the pairs $(1,2),(2,3)\cdots(n-1,n)$, consequently. It seems that the most promising system for implementing such entangled states will be to use ion traps [19-22] where some ions are located on a certain line and the entanglement is produced between neighboring *qubits*, consequently.

Although we have analyzed explicitly the properties of the $3-qubit$ entangled Bell states we get similar properties for Bell states for larger $n-qubit$ systems (n>3). I find that by tracing any Bell state of $n-qubit$ system, over any $(n-2)\,qubits$ of this system, we get $2-qubits$ mixed state which is separable with zero concurrence. The validity of this assumption can easily be checked, for example, by tracing the density matrix of the state $|B'1\rangle$ of (11) over $qubits\ C, D$ obtaining, by using the above methods, mixed density matrix of the $qubits\ A, B$ which is separable and has zero concurrence. Due to symmetry properties of the Bell states such properties are quite general for these states.

In the above analysis we have related certain properties of the general entangled Bell states to concurrence of the mixed states obtained by tracing over certain $qubits$ from the pure general Bell states. Such properties have been found to be in agreement with Peres-Horodecki criterion. We can, however, relate the pure Bell general states to maximal entanglement property in the following way.

Since the entanglement of the general Bell states is produced by consequent entanglement by the Braid operator, operating on pairs $(1,2),(2,3)\cdots(n-1,n)$, we can also study the entanglement property by using the spin-flip operator, operating consequently on the pairs $(1,2),(2,3)$ etc. Then for the pure $3-qubits$ Bell entangled state we can use a generalization of (12) by defining

$$|\tilde{\Phi}\rangle = \left[(\sigma_y \otimes \sigma_y) \otimes I\right] \cdot \left[I \otimes (\sigma_y \otimes \sigma_y)\right] |\Phi^*\rangle \qquad (21)$$

Here $\Phi^*$ is the complex conjugate of $\Phi$ in the standard basis $\{000,001,010,011,100,101,110,111\}$. By taking $|\Phi\rangle$ as any vector corresponding to the state $|Bi\rangle\ (i=1,2,\cdots,8)$ of (10) then we find

$$C(|\Phi\rangle) = |\langle\Phi|\tilde{\Phi}\rangle| = 1 \qquad (22)$$

So, we have shown that the pure $3-qubits$ Bell entangled states have concurrence given by 1, expressing a maximal entanglement. It is quite straight forward to generalize the calculations for large $n-qubits$ Bell states ($n>3$), and find that (22) is satisfied also for these states, so that they have a maximal entanglement.

The main previous works on concurrence have treated bipartite systems [9, 10], or a mixed bipartite system obtained by tracing over one $qubit$ [12]. Therefore, most of the above present analysis for the concurrence of general Bell entangled states, used previous fundamental concurrence properties [9-12]. However, in (21-22) I have generalized the use of spin flip operator, operating consequently on pairs $(1,2),(3,4)$ so by using this new approach I have shown the maximal entanglement of the $3-qubit$ Bell entangled state. An interesting problem arises when we choose $4-qubits$ Bell entangled state, e.g. $|B'1\rangle$ of (11) and we trace its density matrix over $qubit\ D$. In this example we calculate $\rho_{ABC} = Tr_D |B'1\rangle\langle B'1|$. Then by using the generalized spin flip operator I define

$$\tilde{\rho}_{ABC} = \left[\left(\sigma_y \otimes \sigma_y\right) \otimes I\right] \cdot \left[I \otimes \left(\sigma_y \otimes \sigma_y\right)\right] \rho^*_{ABC} \left[\left(\sigma_y \otimes \sigma_y\right) \otimes I\right] \cdot \left[I \otimes \left(\sigma_y \otimes \sigma_y\right)\right] . \qquad (23)$$

By straight forward calculation I find that the square roots of the eigenvalues of $\rho_{ABC}\tilde{\rho}_{ABC}$ are given by $\lambda_1 = \lambda_2 = 1/2$ ; $\lambda_3 = \lambda_4 = \lambda_5 = \lambda_6 = \lambda_7 = \lambda_8 = 0$ , so in analogy to previous calculations the concurrence is vanishing. Due to symmetry properties I find the general result: If in any $4-qubit$ Bell entangled state we trace over one $qubit$ we obtain a mixed state with zero concurrence which includes only classical correlations. In a pictorial description: Assuming that we have $Bell\ entangled\ 4-qubits$ state and that we send the 4 $qubits$ to Alice, Bob, Charles, and Jacob, respectively, which are far, each from the other. If we ignore any measurement made by one of them (e.g. by Jacob) then the other three (e.g. Alice Bob, and Charles) can have only classical correlations between them. Due to the symmetry properties of the general Bell entangled states we can generalize the present properties to any $n-qubit$ entangled Bell states.

6. Summary and conclusions

In the present work certain Braids theories have been developed showing new methods for producing large $n-qubit$ entangled states.

The main Braid operator $R$ [6] is defined in (2). The unitary $R$ operator can be considered in QC as a special universal gate [8] satisfying also the Yang Baxter equation [5-7]. We use for our systems a finite representation for the $Bn$ group [2] given in (5) as $\sigma_1 \sigma_2 \cdots \sigma_{n-1}$. The relation of the $Bn$ operators to the Yang-Baxter equation is described in Figure 1. In QC we use unitary operators where entanglement produced by matrix $R$ can be removed by matrix $R^{-1}$. Such relation is described in Figure 2. In order to get all possible unitary transformations in QC we need to add single *qubits* transformations to the present universal gates.

The present method for producing general entangled $n-qubit$ Bell states ($n>2$) has been analyzed in Section 4. By operating on the computational basis of $n-qubit$ system with the unitary multiplication operator $\sigma_1\sigma_2\cdots\sigma_{n-1}$ orthonormal entangled Bell states are produced. The general entangled $3-qubit$ orthonormal Bell states which can be produced by this method are given in (10). In Figure 3 a general scheme is described for implementing the Braid multiplication $\sigma_1\sigma_2\sigma_3$ which can produce $4-qubit$ orthonormal Bell states, following the present method.

The entanglement properties of the entangled $3-qubit$ Bell states have been related in Section 5 to concurrence [9-12]. In (16-18) we demonstrated a calculation which by tracing over one *qubit* from a certain $3-qubit$ entangled Bell state we obtain a mixed state with zero concurrence, with only classical correlations. A similar conclusion is obtained by checking the Peres-Horodecki criterion, in the same example, using the calculations in (18-19). Due to the symmetry properties of the states of (10) we find that by tracing over any *qubit* from any $3-qubit$ Bell state, we lose all quantum correlations. The implication of such result to information is discussed. In (21-22) I have shown that the $3-qubit$ Bell state has maximal entanglement following from its general concurrence equal 1. We generalize such property to any $n-qubits$ Bell states. By using the general spin-flip operator we find that by tracing the density matrix of $4-qubit$ pure Bell state over one *qubit* we get a mixed state with zero concurrence which can have only classical correlations. These special properties can be generalized to any $n-qubit$ Bell states ($n>2$). The general $n-qubits$ Bell states form a complete set of orthonormal entangled Bell states with maximal entanglement and can be prepared by using a special method.

**Captions for Figures**

Figure1:

The interactions between the three $qubits$ 1,2, and 3 is shown to satisfy the Yang-Baxter equation where the Braid operators $\sigma_1$ and $\sigma_2$ are represented by $R \otimes I$ and $I \otimes R$, respectively, $R$ is given by (1), and $I$ is the $2 \times 2$ unit matrix. The $8 \times 8$ matrix $\sigma_1$ ($\sigma_2$) entangles the $qubits$ 1 and 2 (2 and 3).

Figure 2

A crossing represents the Braid operator $\sigma_i \left( \sigma_i^{-1} \right)$, where a movement, above the crossing , of the right hand from the solid line denoted with an arrow to the other hollow line of the crossing is in opposite clock direction (in the clock direction) . $i$ , $i+1$ represent two neighboring $qubits$ in which the inverse interactions are cancelled, and which are equivalent to $qubits$ which are not interacting, represented by the vertical lines.

Figure 3

The numbers, on the upper horizontal line, denote $qubits$ $1,2,3,4$, respectively. The crossings represent the Braid operators $\sigma_1, \sigma_2$ and $\sigma_3$, given by (5), entangling the $qubits$ pairs $(1,2),(2,3),(3,4)$, consequently. Solid lines connecting neighboring points on vertical lines represent single $qubits$ which are not interacting.